\documentclass[11pt]{article}  

\usepackage{amsmath,amsthm,latexsym,amssymb,amsfonts,epsfig,braket,mathtools,yhmath}

\usepackage{cleveref}

\usepackage{subcaption}
\usepackage{longtable}
\usepackage{booktabs}
\usepackage{fancyvrb}
\usepackage{siunitx}

\usepackage{xcolor}

\usepackage{booktabs}
\usepackage{longtable}
\usepackage{siunitx}

\newtheorem{Definition}{Definition}[section]

\newcommand{\be}{\begin{equation}}
\newcommand{\ee}{\end{equation}}
\newcommand{\ba}{\begin{eqnarray}}
\newcommand{\ea}{\end{eqnarray}}

\title{{\sf Rigorous test of the Raleigh-Ritz method for Mexican hat type potentials}}
\author{
{\sf A. M. Rodriguez Zarate}$^1$\thanks{{\sf 
melissa.rodriguez@gravity.fau.de}}, 
{\sf T. Thiemann}$^1$\thanks{{\sf 
thomas.thiemann@gravity.fau.de}}\\
\\
{\sf $^1$ Inst. for Quantum Gravity, FAU Erlangen -- N\"urnberg,}\\
{\sf Staudtstr. 7, 91058 Erlangen, Germany}\\
}
\date{{\small\sf \today}}

\makeatletter
\@addtoreset{equation}{section}
\makeatother

\begin{document} 

\maketitle

{\sf

\begin{abstract}
Interesting quantum integrable models are rare and one often has to resort to 
approximation methods. One of these is the Raleigh Ritz method which under certain 
circumstances allows to approximately compute the lowest energy eigenstate (or ground state)
of a given Hamiltonian whose pure point spectrum is bounded from below. The quality of such 
approximations can then be tested numerically or sometimes by abstract arguments.

However, the numerical test is limited by computing power. In order to perform a 
rigorous test, one would need to have   
at one's disposal 1. a physically interesting model that is 2. solvable to sufficient 
extent in order that 3. the exact ground state is known in closed form. 

In this contribution we show that certain anharmonic potentials of the Mexican hat 
type belong to this class of models. The corresponding Schr\"odinger type Hamiltonian 
can be considered as a crude quantum mechanical toy model Hamiltonian for the Higgs field in the 
standard model of elementary particle physics.
\end{abstract}

\section{Introduction}
\label{s1}

Apart from a few models of physical interest, most quantum mechanical Hamiltonians
are not quantum integrable in the sense that their (point) spectrum is known 
in closed form. While one can generate a large class of such solvable models by the 
methods of supersymmetric quantum mechanics \cite{1} most of these do not 
model situations of actual physical interest (e.g. molecular Hamiltonians). One therefore 
often has to resort to approximation methods such as WKB methods, perturbation theory 
of point spectra, the Hartree-Fock method, the Raleigh-Ritz method etc. (e.g. 
\cite{2} and references therein). While it is sometimes possible to supply mathematically 
necessary and/or sufficient criteria for the convergence of the corresponding iteration 
methods (e.g. \cite{3} for the perturbation theory of point spectra or \cite{4} for the 
Raleigh-Ritz method), these criteria are typically hard to check and instead one often uses 
numerical convergence tests. Clearly, such numerical methods are limited by the  
computing power of one's machine infrastructure.

The Raleigh-Ritz method (RRM) is a popular tool in order to approximate the (or a, in case of 
degeneracy) ground or vacuum 
state $\Omega$
of a given Hamiltonian $H$ (if it exists). By definition, a ground state is an
eigenstate whose eigenvalue is the lowest in the point spectrum if it is bounded from below
while a vacuum state is an
eigenstate whose eigenvalue is closest to zero in the point spectrum. By shifting 
$H$ by a constant, one can always arrange that zero is in the point spectrum if the 
point spectrum is not empty, but such a corresponding vacuum state may not be a ground state.
On the other hand, a ground state can be arranged to be a vacuum state by shifting $H$ 
by minus the lowest eigenvalue times the unit operator. We will adopt the latter as synonymous:
A vacuum state of a Hamiltonian whose point spectrum is bounded from below is the same as a 
ground state, namely a lowest eigenvalue eigenstate of zero eigenvalue.  

Such a vacuum state is an important state to have at 
one's disposal as it is typically cyclic and thus it tremendously helps to compute 
the matrix elements of the Hamiltonian between any excited states that one obtains 
by acting on it by suitable raising operators.  

The popularity of the RRM rests on the fact that it maps 
a difficult functional analysis problem to a straightforward problem in linear algebra: Given an orthonormal 
basis $b_n,\; n=0, 1, 2,...$ of the (separable) Hilbert space $\cal H$ on which $H$ is a 
self-adjoint operator with dense domain $\cal D$ given by the span of the $b_n$, the procedure 
consists of computing the $N\times N$ Hermitian matrix $H^N$ with entries $<b_m,H\; b_n>,\; 
m,n=0,..,N-1$, diagonalising that matrix and ordering the eigenvalues by size. Let
$\lambda_I^N,\;I=0,..,N-1$ be the eigenvalues of $H_N$ ordered by size 
$\lambda_n^N\le \lambda^N_{n+1}$ (including multiplicity)
and $e^N_I$ the corresponding eigenstates (w.l.g. an orthonormal system). 
Then $e^N_0$ is considered a ground state 
approximant of $\Omega$ with energy lower bound $\lambda^N_0$. In the ideal case one 
would like to show that the sequence of vectors $(e^N_0)_{N\in \mathbb{N}_0}$ is a Cauchy     
sequence but this requires to have access to the $e^N_0$ for all $N$. We refer to 
\cite{4} for criteria that grant that the sequence is indeed Cauchy. In practice 
one is often content with ``numerical evidence'' i.e. one computes the Fourier coefficients 
$c_n^N:=<b_n, e^N_0>,\; n=0,..,N-1$ and shows that for given acceptable error $\epsilon>0$ 
there exists $N_0$ in the computationally accessible range such that $|c_n^N-c_n^{N_0}|<\epsilon$ for all 
$n=0,..,N_0-1$ and all $N>N_0$ in the computationally accessible range $n\le N_M$. 
Here the computational range is defined by the maximal $N_M$ that available computing time allows 
to reach. This is not 
equivalent to $||e^N_0-e^{N_0}_0||< \epsilon$ for some $N_0$ and all $N>N_0$ given $\epsilon$
as this controls also all the Fourier coefficients with index $n>=N_0$. In other words, one shows that 
the $c^N_n$ ``numerically settle'' at given values $c_n$. The convergence rate of the 
procedure of course will depend on the choice of the quite arbitrary ONB $b_n$. If 
$H=H_0+V$ can be split into a solvable part $H_0$ and a perturbation $V$ a physically 
motivated choice is to pick the $b_n$ as eigenstates of $H_0$. 

A model of considerable physical interest intensively studied in the literature using the RRM  
method is the anharmonic oscillator $H=p^2/(2 m)+m\;\omega^2\; q^2/2+g\; q^4/4$ with anharmonic 
$g\;q^4, g>0$ potential contribution and $m,\omega>0$ defining a usual 
harmonic oscillator (see \cite{5} and references therein both for rigorous and numerical 
results). That Hamiltonian can be considered as a toy model for the Higgs particle  
at high temperatures (before spontaneous symmetry breaking). Unfortunately, this Hamiltonian
is not solvable in closed form and in particular the exact ground state $\Omega$ is not 
known in closed form so that one cannot simply compare the $<b_n,e^N_0>$ and $<b_n,\Omega>$.

In this contribution we consider Hamiltonians for which $\Omega$ is known in closed form.
These are constructed by ``backwards engineering''. I.e. instead of providing $H$ and 
computing $\Omega$ we provide $\Omega$ and then compute $H$ such that $H\;\Omega=0$. 
This is inspired by supersymmetric quantum mechanics.
We then tune $\Omega$ such that physically interesting potentials arise. We show that 
we can pick $\Omega$ such that we obtain Hamiltonians with Mexican hat type of potentials
$H=p^2/(2 m)-m\omega^2 q^2/2+g q^{2n}/(2n)$ for some $n\ge 2$ which can be considered as a toy model 
for the Higgs particle at low temperatures (after spontaneous symmetry breaking).
Moreover, one can tune $\Omega$ such that the Fourier coefficients can be computed in closed 
form. Therefore we can rigorously check whether the above ``numerical settlement'' method
can be trusted. \\
\\
This work is organised as follows:\\
\\
In section \ref{s2} we introduce the class of Hamiltonians studied in this work and its 
relation to supersymmetric quantum mechanics. We then establish 
a few of their spectral and computational properties.

In section \ref{s3} we apply the Raleigh - Ritz method with $b_n$ adapted to the harmonic 
part of the potential. We discuss the results of our analysis such as how convergence 
rates depend on the parameters of the potential within the class of chosen potentials. 

In section \ref{s4}
we compare the Raleigh - Ritz method with the more standard quantum perturbation theory
of point spectra.  

In section \ref{s5} we summarise and conclude.

\section{The class of potentials}
\label{s2}

We first recall a few elements from supersymmetric quantum mechanics and then 
introduce the class of superpotentials studied later. We will also establish 
some spectral and computational properties of those potentials.

\subsection{Supersymmetric quantum mechanics}
\label{s2.1}

Throughout this paper we consider the particle on the real line and the Schr\"odinger
representation of the Weyl algebra. Thus the Hilbert space is ${\cal H}=L_2(\mathbb{R},dx)$ 
and position $q$ and momentum operators $p$ respectively are densely defined on Schwarz space 
${\cal D}={\cal S}(\mathbb{R})$ of complex valued smooth functions of rapid decrease at 
infinity. Thus $[q\;\psi](x)=x\;\psi(x),\;[p\;\psi](x)=i\hbar\frac{d}{dx}\psi(x)$ for 
$\psi\in {\cal D}$. 
We are interested in Schr\"odinger type Hamiltonians whose classical symbol
is ($m$ is the mass parameter)
\be \label{2.1}
H=\frac{p^2}{2\;m}+V(q)
\ee
No operator ordering ambiguities arise. 

The class of potentials $V$ that we are interested in are motivated by quantum field theory (QFT):
The Hamiltonian of a real scalar quantum field $\Phi$ in Minkowski spacetime 
$(\mathbb{R}^4,\eta)$ with Minkowski metric $\eta={\sf diag}(-1,1,1,1)$ and 
action $S=\int\;d^4Z\; [-\eta^{\mu\nu}\; \Phi_{,\mu}(Z)\; \Phi_{,\nu}(Z)/2+V(\Phi(Z))]$
is given by 
\be \label{2.2}
H=\int\; d^3z\;[\pi(z)^2/2+\phi(z)[-\Delta\cdot\phi](z)/2+V(\phi(z))]
\ee
where $\phi(z)=\Phi(Z^0=0,\vec{Z}=\vec{z}),\;\pi(z)=[\partial_0 \Phi](Z^0=0,\vec{Z}=\vec{z})$
are the time zero configuration and velocity of the field and $\Delta$ the Laplacian. 
As usual, $H$ arises from $S$ 
by Legendre transformation and $\pi(z),\phi(z)$ obey canonical Poisson brackets 
$\{\pi(z),\phi(z')\}=\delta(z,z')$. In the case that the field $\phi$ and its momentum $\pi$
become spatially homogenous and $\mathbb{R}^3$ is compactified to the torus $T^3$ 
we see that $p=\pi,\; q=\phi$ with $\{p,q\}=1$ and (\ref{2.2}) becomes (\ref{2.1}). Thus 
(\ref{2.1}) is a toy model for the QFT situation.

In the standard model we encounter the Higgs scalar field with Higgs potential $V$. The 
important feature of that potential is that it is a {\bf polynomial} in $q$, typically of fourth 
order, and bounded from below. Therefore we wish to consider precisely such potentials
which are polynomial and and bounded from below, but not necessarily of fourth order.
We now show that this fits into the framework of supersymmetric quantum mechanics. Recall 
that a supersymmetric Hamiltonian is defined by 
\be \label{2.3}
H=A^\dagger\; A,\; A=\frac{\hbar}{\sqrt{2m}}[\frac{d}{dx}-S'(x)]
\ee
where $S$ is a real valued, at least $C^2$ function and we denoted $(.)'=d/dx (.)$. 
It is related to the corresponding 
superpotential $W=-\frac{\hbar}{\sqrt{2m}} \;S'$ and in slight abuse of language we 
call it the superpotential. The corresponding potential 
is given by 
\be \label{2.4}
V=[S']^2+S^{\prime\prime}
\ee
when we work out (\ref{2.3}) to bring it into the form (\ref{2.1}). Here we have 
taken $(\frac{d}{dx})^\dagger=-\frac{d}{dx}$ which needs justification; We assume 
that $H$ is densely defined on a suitable domain $\cal D$ such as ${\cal S}(\mathbb{R})$ 
such that no boundary terms arise when integrating by parts.

A few properties of $H$ can be easily deduced from (\ref{2.3}):\\
1. $H$ is positive and hence symmetric on its domain.\\
2. $\Omega:=e^S$ is a vacuum state if $e^S\in {\cal H}$ (not necessarily normalised).\\
The first property follows from the easy calculation $<\psi,\;H\psi>=||A\psi||^2$ for 
all $\psi\in D$. We take any self-adjoint extension of $H$ granted to exist because 
$H$ is real valued \cite{6}. For the second we first of all have  $A\Omega=0$, hence $H\Omega=0$ 
so that $\Omega$ is a zero eigenvalue eigenstate. Note that an eigenstate must be normalisable,
hence the restriction on $\Omega$. Secondly, from the first property it follows that 
for any eigenstate $\psi$ of $H$ with eigenvalue $\lambda$ we have 
\be \label{2.5}
\lambda=\frac{<\psi,H\;\psi>}{||\psi||^2}\ge 0
\ee
thus the point spectrum of $H$ is bounded from below by zero. Hence $\Omega$ is a vacuum 
state.

Supersymmetric quantum mechanics was invented as a tool to construct Hamiltonians 
of Schr\"odinger type whose point spectrum can be constructed algebraically using raising 
and lowering operators, the prime example being the harmonic oscillator in which case 
$A,A^\dagger$ are simply the familiar annihilation and creation operators. In the course 
of time it transpired that all algebraically solvable potentials are {\bf shape invariant}.
\begin{Definition} \label{def2.1} 
Let $B\subset \mathbb{R}^{M+1},\; M\ge 0$ be a set of parameters $b$. A potential $V_b$ 
deriving from a superpotential $S_b$ depending 
on $b\in B$ is called shape invariant iff there exist functions 
\be \label{2.6}
f:\; B\to B,\; R:\; B\to \mathbb{R}
\ee
such that 
\be \label{2.7}
\tilde{V}_b=V_{f(b)}+R(f(b))\; 1_{{\cal H}}
\ee
where 
\be \label{2.8}
\tilde{V}_b=A_b\; A_b^\dagger-\frac{\hbar^2}{2m}[\frac{d}{dx}]^2
=\frac{\hbar^2}{2m}([S_b']^2-S^{\prime\prime})_b]
\ee
is called the partner potential of $V_b$.
\end{Definition}
Here $A_b=\frac{\hbar}{\sqrt{2m}}(d/dx-S'_b)$ is the corresponding $b$ dependent annihilator
of $\Omega_b=e^{S_b}$. It is not difficult to show that the states $e_{0,b}:=\Omega_b$ and 
for $n\ge 0$
\be \label{2.9}
e_{n+1,b}:=A_b^\dagger\; A_{f(b)}^\dagger\;...\; A_{f^n(b)}^\dagger\; \Omega_{f^{n+1}(b)}
\ee
are eigenstates of $H_b=A_b^\dagger \; A_b$ with eigenvalue $\lambda_{0,b}=0$ and  
\be \label{2.10}
\lambda_{n+1,b}:=\sum_{k=1}^{n+1}\; R(f^n(b))
\ee
respectively as long as (\ref{2.9}) is normalisable. Here $f^n$ is the n-fold application of the map $f$. 
        
As motivated by QFT, we are now asking for superpotentials $S_b$ such that $\Omega_b=e^{S_b}$ is  
normalisable and $V_b$ is a polynomial bounded from below. Evidently this requires 
$[S'_b]^2+S^{\prime\prime}_b=:P_b$ to be a polynomial $P_b$. Substituting $S_b=\ln(\Omega_b)$ this 
Riccati equation becomes the second order linear ODE $\Omega_b^{\prime\prime}=P_b\; \Omega_b$.  
We will not consider the most general solution of this condition (i.e. solving $S_b$ for given 
$P_b$) but simply note that the Riccati condition is obviously satisfied if $S_b$ is itself a 
polynomial
\be \label{2.11} 
S_b=\sum_{m=0}^M\; b_m\; x^m 
\ee
where the dimension of the parameter space minus one $M$ is identified as the polynomial 
degree. The coefficient $b_0$ can be absorbed into the normalisation of $\Omega_b$. 
In order that $\Omega_b$ be normalisable, we must have that $M\ge 2$ is even and that 
the top degree coefficient is negative $b_M<0$. The potential $V_b$ is then also bounded 
from below because $[S'_b]^2=M\; b_M^2\; x^{2[M-1]}+O(x^{2M-3})$ 
while $S^{\prime\prime}_b=M(M-1) b_M\; x^{M-2}$, hence $V_b\ge 0$ for sufficiently 
large $|x|$ and is continuous in between. Thus the parameter space of interest ist 
$B=[\mathbb{R}_+ -\{0\}]\times \mathbb{R}^M$.  

Unfortunately, there is no shape invariant potential in this class except for $M=2$. 
To see this we consider the shape invariance condition 
\be \label{2.12}
[S'_b]^2-S^{\prime\prime}_b=
[S'_{f(b)}]^2-S^{\prime\prime}_{f(b)}+R(f(b))
\ee
This has to hold for all $x\in \mathbb{R}$ and thus is a system of $2(M-1)+1$ equations 
for $M+1$ functions $f_m(b),\; m=1,..,M$ and $R(f(b))$ (the l.h.s. des not depend on 
$b_0$ and the r.h.s. not on $f_0$ if $R$ does not). Thus the number of conditions 
exceeds the number of free functions except when $M=2$ which however leads back 
to the harmonic potential that we are not interested in.

Also, there is no superpotential in this class such that $V_b$ has degree four. 
The lowest possible degree in this class such that $V_b$ is not harmonic 
is six for the case $M=2$. However, for any superpotential in this class the vector $\Omega_b$ 
is cyclic. For $M=2$ this is trivial as the $q^n \Omega_b,\; n\in \mathbb{N}_0$ 
exhaust all Hermite functions which lie dense. For $M>2$ even the $q^n \Omega_b$ lie also dense 
for suppose there exists $\psi$ orthogonal to the closure of this span. Pick any $c>0$ then 
for any $n$
\be \label{2.13}
0=<\psi, q^n\; \Omega_b>=<\Omega_b(q)\; e^{c q^2}\;\psi,\; q^n \; e^{-c q^2}\; 1>
\ee
The operator $\Omega_b(q)\; e^{c\; q^2}$ acts by multiplication by the bounded function 
$\Omega_b(x)\; e^{c\; x^2}$ hence $\psi':=\Omega_b(q)\; e^{c\; q^2}\psi\in {\cal H}$. The    
r.h.s. of (\ref{2.13}) exhausts all Hermite functions, hence $\psi'=0$ i.e. 
$\psi'(x)$ a.e. with respect to $dx$. Since $\Omega_b(x)\; e^{c\; x^2}$ is nowhere vanishing
it follows that $\psi(x)=0$ a.e. hence $\psi=0$. 

\subsection{Chosen set of superpotentials}
\label{s2.2}

In what follows we will restrict ourselves to the two parameter class of super potentials
\be \label{2.14}
S_b(x)=\sigma\; \frac{x^2}{2\;l^2}-\frac{x^{2k}}{(2k)\;L^{2k}}
\ee
where $l, L>0$ have dimension of length, $k\ge 2$ and $\sigma=\pm 1$. The corresponding
potential is given by 
\be \label{2.15}
V_b(x)
=\sigma\frac{1}{l^2}+\frac{1}{l^4}\;x^2-(2k-1)\frac{x^{2(k-1)}}{L^{2k}}
-2\sigma \frac{1}{l^2 L^{2k}} x^{2k}+\frac{x^{4k-2}}{L^{4k}}
\ee
It is obviously bounded from below and reflection symmetric. 
It has a zero point energy $V_b(0)\not=0$ and for $k>2$ it always has 
a harmonic term. For $k=2$ it has an anti-harmonic term when $l^{-4}<3 L^{-4}$. For all $k\ge 2$ 
it has the anharmonic term of top degree $4k-2$. For $k=2$ and $\sigma=-1$ and $l^{-4}\ge 3 \;L^{-4}$
the potential $V_b+l^{-2}$ is manifestly not negative, otherwise it may take also negative values depending 
on the ratio $\delta:=[l/L]^{2k}$. Nevertheless, the energy spectrum of the 
Hamiltonian is not negative for all parameters by design. We see that 
for the case $k=2$ the sixtic potential shares with quartic Higgs potential the feature that 
for $\delta >\frac{1}{3}$ we obtain a negative mass squared term (low temperature phase) 
which gives rise to local minima away from zero (condensates) while 
for  $\delta <\frac{1}{3}$ we obtain a positive mass squared term (high temperature phase).
The feature of obtaining local minima different from zero is also true for all $k>=2$ when 
$\sigma=1$ no matter what the value of $\delta$ is. Thus in what follows we will consider 
the case $\sigma=1$ as a model for the low temperature Higgs field. To see this in an example 
consider the case $k=2,\sigma=1$. Then in terms of the dimension free variable $z=y^2,\; 
y=x/l$
\be \label{2.14}
F(z):=l^2 V_b(x)
=1+[1-3\delta]\; z
-2\delta\; z^2+\delta^2\;z^3
\ee 
Hence 
\be \label{2.17}
F'(z)
=3\delta^2\;[z-\frac{1}{3\delta}(2+\sqrt{1+9\delta})]\;[z-\frac{1}{3\delta}(2-\sqrt{1+9\delta})]
\ee
For $\delta>1/3$ this vanishes only at $z=z_-=\frac{1}{3\delta}(2+\sqrt{1+9\delta})$ while 
for $\delta<1/3$ this vanishes also at $z=z_+=\frac{1}{3\delta}(2-\sqrt{1+9\delta})$. In the first 
case we have $F'<0$ for $0<z<z_-$ and $F'>0$ for $z>z_-$.  In the second
case we have $F'>0$ for $0<z<z_+, z>z_-$ and $F'<0$ for $z_+<z<z_-$. Thus in both cases $z_-$ is a local 
minimum. Similar considerations hold for $k>2$ except that in this case one can no longer 
locate the minimum algebraically for $k>4$ and generic values of $\delta$.\\
\\
Further properties of $H$ are as follows:\\ 
Clearly, since $H$ is a positive, hence 
symmetric operator on the domain of Schwarz functions we can take the Friedrichs
self-adjoint extension \cite{4}. However, by the results of \cite{7}, $H$ is even essentially 
self-adjoint on the domain of smooth functions with compact support and its essential 
spectrum is empty, i.e. its spectrum is purely discrete consisting only of isolated 
eigenvalues with finite multiplicity. Furthermore by the minimax principle \cite{4},
given any $N-$dimensional subspace, the $N$ eigenvalues of the projection $H_N$ to that space
ordered by size provide upper bounds to the first $N$ eigenvalues of $H$ ordered by size.

\subsection{Exact ground states}
\label{s2.3}

With regard to the concrete application of the RRM, we must pick an ONB with respect to which we compute the matrix 
elements. We will choose the $b_n$ to be the Hermite functions of the harmonic 
oscillator with length parameter $l^2=\hbar/(m\omega)$. 
Moreover, in this subsection we will use $\epsilon=l/ L$ instead of $
\delta=\epsilon^{2k}=[l/L]^{2k}$ of the previous subsection.
Thus we have the usual energy eigenfunctions of the quantum harmonic oscillator
\begin{align}
    b_n(x) = \frac{\pi^{-1/4}}{\sqrt{2^n n!}}e^{\frac{-x^2}{2}}H_{n}(x),
\end{align}
where $H_{n}$ corresponds to the physicist's Hermite polynomials which can be defined in terms of its even/odd series expansion
\begin{equation}
\label{hermite}
H_{n}(x)=\left\{ \begin{aligned} 
n!\sum_{l=0}^{n/2}\frac{(-1)^{\frac{n}{2}-l}(2x)^{2l}}{(2l)!(\frac{n}{2}-l)!}\hspace{5mm}\text{for even n}, \\
   n!\sum_{l=0}^{\frac{n-1}{2}}\frac{(-1)^{\frac{n-1}{2}-l}(2x)^{2l+1}}{(2l+1)!(\frac{n-1}{2}-l)!}\hspace{5mm}\text{for odd n}. 
\end{aligned} \right.
\end{equation}
Hence the Fourier coefficients with respect to the energy eigenbasis for even $n$ are given by
\begin{align}
\label{exact}
I_n:=\braket{b_n,\Omega_b}  & = \frac{\pi^{-1/4}}{\sqrt{2^n n!}}\int_{-\infty}^{\infty} H_{n}(x)e^{-\frac{\epsilon^{2k}x^{2k}}{2k}}dx \nonumber \\ 
    & = \frac{\pi^{-1/4}n!}{\sqrt{2^n n! }} \sum_{m=0}^{n/2}\frac{(-1)^{\frac{n}{2}-m}2^{2m+1}}{(2m)!(\frac{n}{2}-m)!}\int_{0}^{\infty} x^{2m}e^{-\frac{\epsilon^{2k}x^{2k}}{2k}}dx \nonumber \\ &  =\frac{\pi^{-1/4}n!}{\sqrt{2^n n!}} \sum_{m=0}^{n/2}\frac{(-1)^{\frac{n}{2}-m}2^{\frac{2k2m+2m+1}{2k}}k^{\frac{2m-2k+1}{2k}}}{(2m)!(\frac{n}{2}-m)!\epsilon^{2m+1}}\Gamma\left(\frac{2m+1}{2k}\right),
\end{align}
where in the second line we used the definition of the Hermite polynomials for even $n$ and we solved the integral by the change of of variables
$u=\frac{\epsilon^{2k}x^{2k}}{2k}$. The Fourier coefficients for $n$ odd trivially vanish as $\Omega_b, \; H_n$ are even and odd 
functions respectively under reflection.  

We also note that for any choice of polynomial superpotential
$S_b$ the computation
of $I_n=<b_n,\Omega>$  can be reduced to that of the
$J_n=\int\;dy\;y^n\;e^{-y^2/2+S_b(y)}$ in terms of the dimensionfree
variable $y=x/l$. Now we have the identity
$-n\;J_{n-1}=\int\;dy\;y^n\;[-y+S'_b(y)]\;e^{-y^2/2+S_b(y)}$ by the
properties of $S_b$ and integration by parts.
If $S_b$ is a polynomial of degree $M$ then
the right hand side of this identity is a linear combination of
$J_m$ with $m-n=0,1,2..,M-1$. Thus $J_n$ is a known linear combination of the
$J_{n-1},..,J_{n-M}$ and thus only the $J_0,..,J_{M-1}$ need to be known
numerically, all others can be computed algebraically using
this recursion relation.
For our class of $S_b$ we have $-y+S'_b=-\delta\; y^{2k-1}$,
thus the recursion becomes especially simple
$n\;J_{n-1}=\delta \; J_{n-1+2k}$ which was part of the motivation for
their choice. Of course, this is also directly reflected in (\ref{exact})
in the corresponding recursion $\Gamma(z+1)=\Gamma(z)$ for the $\Gamma$
function.

An issue to be careful about when applying the RRM is that the exact ground state $\Omega$ corresponding to 
(\ref{2.14}) is not normalised and that the convergent integral
\be \label{2.19}
||\Omega||^2=\int_{\infty}^\infty\; dx \; e^{2\;S_b(x)}=\int_{-\infty}^{\infty}e^{-\frac{\epsilon^{2k}x^{2k}}{k}+x^2} dx,
\ee
is not expressible in terms on known functions for general $k$. 
However it turns out that for the specific models studied below (for $k=2, 3$ and $4$) the normalisation factor can be written
in terms of known functions. 
The computation of their analytic expression was performed directly in \texttt{Mathematica 14.1} using the \texttt{Integrate} command. 
The expressions are far from trivial and are given in terms of Modified Bessel functions of the first kind, Airy functions and 
generalised hypergeometric functions for $k=2,3,4$ respectively.
For the model $k=2$ the exact normalisation factor is 
\begin{align}
\label{k=2}
  \int_{-\infty}^{\infty}e^{-\frac{\epsilon^{4}x^{4}}{2}+x^2} dx=\frac{\pi}{2 \epsilon ^2} exp[1/4 \epsilon ^4] \left[I_{\frac{1}{4}}\left(\frac{1}{4 \epsilon ^4}\right)+I_{-\frac{1}{4}}\left(\frac{1}{4 \epsilon ^4}\right)\right],
\end{align}
where $I_\nu[z]$ corresponds to the Modified Bessel function of the first kind and $\epsilon$ is assumed to be positive real. 
To deduce \cref{k=2} one may rewrite the integral as \(\int_0^\infty exp[-\mu x^4-2\nu x^2]dx\), relate it to the Modified Bessel functions of the second kind $K_\nu[z]$ and then use the known identities between $I_\nu[z]$ and $K_\nu[z]$. 
Alternatively one may perform a change of variables to rewrite it as \(\int_0^\infty x^{\nu-1}exp[-\beta x^2-\gamma x]dx\) which is related to parabolic cylinder functions $D_\nu[z]$ and then use known identities between $D_\nu[z]$ and $I_\nu[z]$. 
These -and related- integrals can be found in \cite{8,9}. 
For k=3 the normalisation factor is
\begin{align}
\label{k=3}
  \int_{-\infty}^{\infty}e^{-\frac{\epsilon^{6}x^{6}}{3}+x^2} dx=\frac{\pi ^{3/2}}{2^{1/3}\epsilon} \left[\text{Ai}\left(\frac{1}{2^{2/3} \epsilon ^2}\right)^2+\text{Bi}\left(\frac{1}{2^{2/3} \epsilon ^2}\right)^2\right],
  \end{align}
where Ai[z] and Bi[z] correspond to the Airy functions.
Equation (\ref{k=3}) is obtained after performing a change of variables to relate it to the integral representation of Airy functions \(\text{Ai}^2(z)+\text{Bi}^2(z)=\frac{1}{\pi^{3/2}}\int_0^\infty exp[zx-\frac{x^3}{12}]x^{-1/2}dx\),
see \cite{9} and references therein.
Finally for k=4 we have
\begin{align}
\label{k=4}
  \int_{-\infty}^{\infty}e^{-\frac{\epsilon^{8}x^{8}}{4}+x^2} dx&=\frac{\pi ^{3/2}}{128\sqrt{2}\;\epsilon^7} \left[\frac{128\epsilon ^6 \, _1F_3\left(\frac{5}{8};\frac{3}{4},\frac{5}{4},\frac{3}{2};\frac{1}{64 \epsilon ^8}\right)}{\Gamma \left(\frac{1}{8}\right) \Gamma \left(\frac{3}{4}\right)}+\frac{128 \epsilon^6 \, _1F_3\left(\frac{1}{8};\frac{1}{4},\frac{1}{2},\frac{3}{4};\frac{1}{64 \epsilon ^8}\right)}{\Gamma \left(\frac{5}{8}\right) \Gamma \left(\frac{3}{4}\right)}+\right. \nonumber \\ & \left.\frac{5 \, _1F_3\left(\frac{7}{8};\frac{5}{4},\frac{3}{2},\frac{7}{4};\frac{1}{64 \epsilon ^8}\right)}{\Gamma \left(\frac{11}{8}\right) \Gamma \left(\frac{9}{4}\right)}-\frac{256 \sqrt{\epsilon ^8} \, _1F_3\left(\frac{3}{8};\frac{1}{2},\frac{3}{4},\frac{5}{4};\frac{1}{64 \epsilon ^8}\right)}{\Gamma \left(-\frac{1}{8}\right) \Gamma \left(\frac{5}{4}\right)}\right],
  \end{align}
where $_1F_3$ are Generalised Hypergeometric Functions.  
To evaluate the integral, one performs a change of variables to cast it into Laplace-type form \cite{10}. 
The factor $e^{x^2}$ is then expanded as a power series, and the sum is interchanged with the integral.  
Although $e^{x^2}$ grows for large $|x|$, the dominant damping from $e^{-\varepsilon^8 x^8}$ ensures convergence of each term.  
This allows the integral to be computed term by term, yielding expressions involving Gamma functions. Mathematica then identifies the resulting series as a linear combination of generalized hypergeometric functions. 
Here we only sketched the proofs since the Mathematica \texttt{Integrate} command is highly reliable, 
furthermore one can verify the correctness of that symbolic result by numerically integrating using \texttt{NIntegrate} 
and comparing to the closed-form \cref{k=2,k=3,k=4}.
In the following table we show the error estimate between the closed form expression obtained by applying 
\texttt{Integrate} and by applying numerical integration \texttt{NIntegrate} respectively for $\epsilon=2$. 
The latter command applies a highly adaptive numerical integration procedure that automatically chooses the most 
suitable integration method based on the chosen integrand and the tuneable options that one may select. 
With the aid of those options one may control which approximation method is used. We tabulate that error both 
for the fully adaptive method (all options default) and by picking options to select the so called trapezoidal method. 
As it can be seen, the numerical agreement confirms the validity of the symbolic expression produced by \texttt{Mathematica} for 
positive real $\epsilon$.
 \begin{longtable}{ccc}
 \toprule
 k & \text{Default} 
 & \text{Trapezoidal} \\
 \midrule
2 & 7.618350394977824e-13 & 1.5324852498110886e-11 \\
 3 & 7.086109476972524e-12 & 1.7763568394002505e-15 \\
 4 & 3.503863865716994e-13 & 0. \\
 \bottomrule
 \caption{Error estimates for $k=2,3,4$ and $\epsilon=2$ for the default approximation method of \texttt{NIntegrate} and the chosen trapezoidal method.}
\end{longtable}  

\section{Application of the Raleigh-Ritz Method}
\label{s3}

In this section we employ the RRM to find the ground state and its energy eigenvalue of the class of Hamiltonians
introduced in section \ref{s2} parametrised by $k\ge 2,\;\sigma=1$ and $\epsilon$. 
As in \cref{s2.3} we define $\epsilon=l/ L$ instead of $
\epsilon=[l/L]^{2k}$. 
As stated in \cref{s1} to apply the RRM it is customary to find the explicit form of the N-th dimensional Hamiltonian matrix 
$H^N_{mn}:=\braket{b_m,Hb_n},\; m,n\le N$. 
For reasons of being complete we computed the analytic expression for each of the entries of $H_{mn}^N$. 
In practice however, the matrix was directly computed in \texttt{Mathematica} using the \texttt{cuantica} and \texttt{versora} packages \cite{11}. 
These packages are particularly useful to symbolically solve quantum mechanical models. 
In particular we used their build-in creation and annihilation operators to compute the matrix $H^N_{mn}$. 
Further details on the code can be found in \cref{a2}.
It is convenient to compute each term of the Hamiltonian matrix separately. 
The free part as well as the $\frac{1}{2}$ factor have a trivial matrix form.  
For the 
$x^{2k}$ term we have
\begin{align}
\label{69}
    <b_m,x^{2k}b_n> & =\int_{-\infty}^{\infty}\frac{\pi^{-1/2}e^{-x^2}x^{2k}}{\sqrt{2^{m+n}n!m!}}H_{m}(x)H_{n}(x)dx.
\end{align}
The Hermite polynomials given in \cref{hermite} can also be combined using floor functions
\begin{gather}
\label{hermitefloor}
H_m(x)=m!\sum_{j=0}^{\lfloor m/2 \rfloor}\frac{(-1)^{j}(2x)^{m-2j}}{j!(m-2j)!}.
\end{gather}
By means of \cref{hermitefloor} we compute the integral
\begin{align}
  \int_{-\infty}^{\infty}dx\hspace{1mm}e^{-x^2}x^{2k}H_{m}(x)H_{n}(x) & = m!n!\sum_{j=0}^{\lfloor m/2 \rfloor}\sum_{i=0}^{\lfloor n/2 \rfloor}\frac{(-1)^{i+j}2^{m+n-2(i+j)}}{i!j!(m-2j)!(n-2i)!}\int_{-\infty}^{\infty}e^{-x^2}x^{2(k-i-j)+m+n}dx \nonumber \\
  & = m!n!\sum_{j=0}^{\lfloor m/2 \rfloor}\sum_{i=0}^{\lfloor n/2 \rfloor}\frac{(-1)^{i+j}2^{m+n+2(k-i-j)}}{i!j!(m-2j)!(n-2i)!}  \Gamma \left[\frac{ (-2 i-2 j+2 k+m+n+1)}{2}\right]
\end{align}
which only holds for $m$ and $n$ of the same parity. 
When they have different parity, the \(x^{2(k-i-j)+m+n}\) term is odd and thus the whole integral is odd.
Thus, we only have contributions when both $m$ and $n$ have the same parity. 
The same computations can be performed for the other polynomial terms in \cref{2.15}.
All other terms yield to the same integral form and therefore can be written in terms of Gamma functions.
However as stated, the matrix was directly computed by \texttt{Mathematica} using the code in \cref{a2}. \\
\\
In what follows we report the RRM results for both the ground state energy eigenvalue and the 
ground state Fourier coefficients with respect to the harmonic oscillator orthonormal basis.

\subsection{Ground state energy eigenvalue}
\label{energies}

In the following we show some of the results obtained for different combinations of the coupling parameters $\epsilon$ and the polynomial degrees $k$. 
Tables 2. to 6. show the values for the ground state energy for fixed $\epsilon, \;k$ and increasing dimension $N$.
As depicted in the tables, for fixed $\epsilon$, convergence becomes more difficult as $k$ increases.
Likewise, for fixed $k$, convergence becomes more difficult as $\epsilon$ increases. 
Larger matrices ($N > 350$) were not investigated due to memory limitations. 
Additionally, values of $\epsilon > 2$ and $k > 5$ proved to be computationally expensive even for relatively small matrix sizes.
Meaningful convergence in those cases would require very large matrices which were not accessible. 
In these regimes, meaningful convergence appears to require significantly larger matrices, which exceeded the available computational resources.
All computations were performed on a personal workstation without the use of institutional servers or parallelised kernels.

\begin{table}[ht]
\centering
\begin{minipage}{0.48\textwidth}
\centering
\begin{tabular}{c S}
\toprule
N & {Eigenvalue} \\
\midrule
25  & 0.53714608101508610000 \\
50  & 0.05262033050727558000 \\
75  & 0.02599471478885967000 \\
100 & 0.00203148697289384600 \\
125 & 0.00130912318171962500 \\
150 & 0.00014460056441802190 \\
200 & 0.00002038390838210715 \\
300 & 8.84519257939422935219e-8 \\
\bottomrule
\end{tabular}
\caption{Ground state eigenvalues for $\epsilon = 2$, $k = 2$.}
\end{minipage}
\hfill
\begin{minipage}{0.48\textwidth}
\centering
\begin{tabular}{c S}
\toprule
N & {Eigenvalue} \\
\midrule
25 & 9.196381553950952 \\
 100 & 0.5153238893959655 \\
 200 & 0.07671396814916273 \\
 300 & 0.01012446872676996  \\
\bottomrule
\end{tabular}
\caption{Ground state eigenvalues for $\epsilon = 2$, $k = 3$.}
\end{minipage}
\end{table}

\begin{table}[ht]
\centering
\begin{minipage}{0.48\textwidth}
\centering
\begin{tabular}{c S}
\toprule
N & {Eigenvalue} \\
\midrule
 100 & 1.8789460770708353703989608381e-9 \\
 125 & 7.781837542781401708593575033e-11 \\
 150 & 2.94458191583850456772516458e-12 \\
 175 & 1.1839385585495514329714784e-13 \\
 200 & 4.413388536743133945688178444653867177560917e-15 \\
 300 & -3.245232940655732914463251541929146267581e-19 \\
 325 & -2.086594795718469213003613686824441969744e-19 \\
 350 & 2.343872379885173983454981756909001849114e-19 \\
\bottomrule
\end{tabular}
\caption{Ground state eigenvalues for $\epsilon = 1$, $k = 2$.}
\end{minipage}
\hfill
\begin{minipage}{0.48\textwidth}
\centering
\begin{tabular}{c S}
\toprule
N & {Eigenvalue} \\
\midrule
 100 & 0.0008728751133444515 \\
 125 & 0.0002344626774069086 \\
 150 & 0.00007126823844764218 \\
 175 & 0.00002176942921548070 \\
 200 & 3.84389957455671961003185041732783e-6 \\
 225 & 1.33592603977971869716363401292228e-6 \\
 250 & 5.3173784768747485637613593863096e-7 \\
 300 & 4.944243834708156999344192700485e-8 \\
\bottomrule
\end{tabular}
\caption{Ground state eigenvalues for $\epsilon = 1$, $k = 3$.}
\end{minipage}
\end{table}

\begin{longtable}{c S S S}
\toprule
N & {Eigenvalue for $k=2$} & {Eigenvalue for $k=3$} & {Eigenvalue for $k=4$} \\
\midrule
\endfirsthead
\toprule
N & {Eigenvalue for $k=2$} & {Eigenvalue for $k=3$} & {Eigenvalue for $k=4$} \\
\midrule
\endhead
25  & 0.9815159620090416       & 0.06494556855120519      & 0.19223341842217834       \\
50  & 0.14442517073412786      & 5.6938036335497e-5       & 0.003870401043707188      \\
75  & 1.4418086714811137e-8    & 1.2115081964752837e-7    & 0.00010451649017137993    \\
100 & 7.132035697334496e-15    & 3.391620141188376e-10    & 3.1282262685617844e-6     \\
125 & 2.7330987307456215e-15   & 2.9573605092491783e-12   & 2.3970039966194426e-7     \\
150 & 9.67536416102255e-15     & 2.1656581258796848e-14   & 1.1002577908945878e-8     \\
\bottomrule
\caption{Ground state eigenvalues for $\epsilon = 0.33$ and $k = 2$, $3$, and $4$.}
\end{longtable}

\subsection{Ground state Fourier coefficients}
\label{eigenstates}

The main aim in this section is to compare the exact Fourier coefficients $\frac{\braket{b_n,\Omega}}{||\Omega||}$ 
from \cref{s2.3} with the coefficients  $\braket{b_n, e^N_0}$ obtained by the RRM.
Note that by construction the exact $\frac{\braket{b_n,\Omega}}{||\Omega||}$ is available analytically, the 
$<b_n,e^0_N>$ is only known numerically simply because the eigenvalue problem for an N times N matrix 
can only be solved numerically. 
We plugged the functions (\ref{exact}) with their respective normalisations \cref{k=2,k=3,k=4} in \texttt{Mathematica} and call them \texttt{omegaexact}.
Then, for fixed $\epsilon$ and $k$ we computed numerical tables of the form
\begin{verbatim}
omegaexact = Block[{$MaxExtraPrecision = 1000},N[omega[#] & /@ Table[i, {i, 0, N-1, 2}], 
            {\[Infinity], 16}]];
\end{verbatim}
\texttt{\$MaxExtraPrecision=1000} is used to minimize roundoff errors in intermediate computations within \texttt{Block} and we chose the target output precision to be 16 digits.
The table runs up to the dimension of the orthonormal basis and takes steps of two since -due to \cref{exact}- the odd entries are zero.

For the numerical approximation using the RRM we ran the code given in \cref{a2}.
The \texttt{Eigensystem} command provided us with an orthonormal system. 
The Fourier coefficients $\braket{b_n, e^N_0}$ correspond to the entries of the eigenvector associated to the ground state energy.

In the following tables we fixed $\epsilon$ and $k$ and computed the exact and approximated coefficients up to some truncated dimension $N$.
The last column reports the estimated error, given by the absolute value of the difference between the exact and approximated results.
More coefficients imply more numerical precision, for that reason we considered only the range between $N\geq 100$
and $N\leq 300$ since the computations for higher $N$'s are computationally expensive and for lower $N$ not accurate enough. 
The tables for the non vanishing coefficients $N \in [100,300]$ would have rows $N/2$, for reasons of space
we confined the tables to $\approx$ 16 coefficients.
In practice however, all the coefficients where computed. To check the error on the totality of coefficients 
computed we also calculated the approximate norms squared (i.e. the sum of moduli squared of the coefficients up to the 
given value $N$)for both the exact and numerical coefficients.
The approximate norms in both cases should be very close to and bounded from above by 1 by Bessel's inequality
since the coefficients are with respect to an orthonormal basis.

As in \cref{energies}, for fixed $\epsilon$, convergence becomes more difficult as $k$ increases.
Likewise, for fixed $k$, convergence becomes more difficult as $\epsilon$ increases. 
Additionally, values of $\epsilon > 2$ and $k > 3$ proved to be computationally very expensive even for relatively small $N$, where the error estimates are considerably large.
In the case of the exact coefficients, $0<\epsilon < 1$ proved to be numerically challenging since the $\epsilon$'s in the denominator of \cref{exact} are very small and Mathematica interprets them as dividing by zero.
This could not be resolved by \texttt{\$MaxExtraPrecision}.

\begin{longtable}{S S S}
\toprule
{Exact} & {Approximated} & {Error} \\
\midrule
\endfirsthead
\toprule
{Exact} & {Approximated} & {Error} \\
\midrule
\endhead
 0.943478717888305 & 0.943478718464222 & 5.759173369133639581727181471e-10 \\
 0.234804316310354 & 0.2348043150452623 & 1.2650912854820408912808165852e-9 \\
 -0.214106218852537 & -0.2141062182522063 & 6.0033092371718049679885049703e-10 \\
 0.072485857223595 & 0.0724858575248454 & 3.0125032407139801784537781476e-10 \\
 0.013724252087633 & 0.01372425114407816 & 9.4355443735522318819052073884e-10 \\
 -0.040591775201398 & -0.04059177411878774 & 1.08261018355335522749553951965e-9 \\
 0.035137177765054 & 0.03513717708868294 & 6.76371401767583913450111068e-10 \\
 -0.019461411685357 & -0.01946141182485428 & 1.3949772788013185472156652209e-10 \\
 0.005274536469501 & 0.00527453756470205 & 1.09520097777287381919055150878e-9 \\
 0.003454483598383 & 0.003454481742431781 & 1.85595093661417531651945906494e-9 \\
 -0.006897671114156 & -0.00689766900656456 & 2.10759121796678878211400238446e-9 \\
 0.006768251813932 & 0.00676825017138583 & 1.6425457366859132351548508953e-9 \\
 -0.004890075474266 & -0.00489007504625973 & 4.2800653482159639554888851194e-10 \\
 0.002610311003121 & 0.002610312362686764 & 1.35956587404041122047013661041e-9 \\
 -0.000693972557276 & -0.000693975901385243 & 3.34410942482713153472538496452e-9 \\
\bottomrule
\caption{List of coefficients for $\epsilon=1$, $k=2$, $N=100$. Only the first 15 coefficients are shown.}
\end{longtable}

\begin{longtable}{S S S}
\toprule
{Exact} & {Approximated} & {Error} \\
\midrule
\endfirsthead
\toprule
{Exact} & {Approximated} & {Error} \\
\midrule
\endhead
 0.943478717888305 & 0.943478717889220 & 9.1552871236182308313221789e-13 \\
 0.234804316310354 & 0.2348043163084072 & 1.94635010312902601068730994e-12 \\
 -0.214106218852537 & -0.2141062188515893 & 9.478871428117962286128919e-13 \\
 0.072485857223595 & 0.0724858572240324 & 4.373085821716242872025717e-13 \\
 0.013724252087633 & 0.01372425208618493 & 1.44767321505900442144116788e-12 \\
 -0.040591775201398 & -0.04059177519970540 & 1.69251751749767332260475423e-12 \\
 0.035137177765054 & 0.03513717776396405 & 1.09028702264972093858181513e-12 \\
 -0.019461411685357 & -0.01946141168551885 & 1.6230024170992886564896606e-13 \\
 0.005274536469501 & 0.00527453647115554 & 1.65447622760821835500337832e-12 \\
 0.003454483598383 & 0.003454483595515588 & 2.8671288148831391027816584e-12 \\
 -0.006897671114156 & -0.00689767111085013 & 3.30564631546686833433516179e-12 \\
 0.006768251813932 & 0.00676825181129677 & 2.6348039060737969296516147e-12 \\
 -0.004890075474266 & -0.00489007547347882 & 7.8744676285117695437227027e-13 \\
 0.002610311003121 & 0.002610311005099064 & 1.97817380127431309004454918e-12 \\
 -0.000693972557276 & -0.000693972562365150 & 5.08933137976965269273967587e-12 \\
 -0.000564556872436 & -0.000564556864692479 & 7.74308275245499864271478839e-12 \\
\bottomrule
\caption{List of coefficients for $\epsilon=1$, $k=2$ and $N=150$}
\end{longtable}

\begin{longtable}{S S S}
\toprule
{Exact} & {Approximated} & {Error} \\
\midrule
\endfirsthead
\toprule
{Exact} & {Approximated} & {Error} \\
\midrule
\endhead
 0.943478717888305 & 0.943478717888306 & 1.39599203713902685417969e-15 \\
 0.234804316310354 & 0.2348043163103508 & 2.7914469773403562662102e-15 \\
 -0.214106218852537 & -0.2141062188525358 & 1.44954224642154387725249e-15 \\
 0.072485857223595 & 0.0724858572235956 & 5.2914719628245779927865e-16 \\
 0.013724252087633 & 0.01372425208763054 & 2.05890283048504299523601e-15 \\
 -0.040591775201398 & -0.04059177520139540 & 2.52440104101598142704767e-15 \\
 0.035137177765054 & 0.03513717776505259 & 1.7468860279593203678619e-15 \\
 -0.019461411685357 & -0.01946141168535659 & 
0.0 \\
 0.005274536469501 & 0.00527453646950332 & 2.25331539564784696004623e-15 \\
 0.003454483598383 & 0.003454483598378571 & 4.14603254139058669938017e-15 \\
 -0.006897671114156 & -0.00689767111415081 & 4.96478042741769482640265e-15 \\
 0.006768251813932 & 0.00676825181392740 & 4.17088105641532108118572e-15 \\
 -0.004890075474266 & -0.00489007547426466 & 1.60618201414797174273527e-15 \\
 0.002610311003121 & 0.002610311003123299 & 2.4091248223401322579081e-15 \\
 -0.000693972557276 & -0.000693972557282896 & 7.07750422940481824634805e-15 \\
 \bottomrule
 \caption{List of first coefficients for $\epsilon=1$, $k=2$ and $N=200$}
 \end{longtable}

\begin{longtable}{S S S}
\toprule
{Exact} & {Approximated} & {Error} \\
\midrule
\endfirsthead
\toprule
{Exact} & {Approximated} & {Error} \\
\midrule
\endhead
 0.939573916378489 & 0.9395855632800647 & 0.000011646901576 \\
 0.104462343230550 & 0.1044410284579323 & 0.000021314772618 \\
 -0.242618145795966 & -0.2426017340805843 & 0.000016411715382 \\
 0.171571988678263 & 0.1715646078354353 & 7.38084282779865103067549910311948e-6 \\
 -0.069276319610354 & -0.06927764543300931 & 1.32582265554427448392888629546702e-6 \\
 -0.008205789264892 & -0.008198208934560699 & 7.58033033167777926136518833261359e-6 \\
 0.050136890718691 & 0.05012626973437683 & 0.000010620984314 \\
 -0.062486711605800 & -0.06247616177864163 & 0.000010549827159 \\
 0.055642809404103 & 0.05563481720865271 & 7.99219545022351320107563332673058e-6 \\
 -0.039269392294065 & -0.03926555041177620 & 3.84188228910932510637326162323921e-6 \\
 0.020536072745919 & 0.02053700634526511 & 9.3359934631530086566917756686691e-7 \\
 -0.003909743322002 & -0.003915181019791956 & 5.43769778962242630264412175759546e-6 \\
 -0.008402891800985 & -0.008393947059482988 & 8.94474150199651751502365226062521e-6 \\
 0.015821326011614 & 0.01581036656598755 & 0.000010959445627 \\
 -0.018783879401313 & -0.01877263861911690 & 0.000011240782196 \\
\bottomrule
\caption{List for first coefficients for $\epsilon=1$, $k=3$ and $N=150$.}
\end{longtable}

\begin{longtable}{S S S}
\toprule
{Exact} & {Approximated} & {Error} \\
\midrule
\endfirsthead
\toprule
{Exact} & {Approximated} & {Error} \\
\midrule
\endhead
 0.939573916378489 & 0.9395739246443276 & 8.26583853473973019463503482225e-9 \\
 0.104462343230550 & 0.1044623284512477 & 1.477930279220705320285242657467e-8 \\
 -0.242618145795966 & -0.2426181344558999 & 1.134006606512764527097146949218e-8 \\
 0.171571988678263 & 0.1715719835990493 & 5.0792137929241002739557823624e-9 \\
 -0.069276319610354 & -0.06927632055493002 & 9.4457626118341868762997846707e-10 \\
 -0.008205789264892 & -0.008205783998466138 & 5.26642623893711357910096697902e-9 \\
 0.050136890718691 & 0.05013688335648565 & 7.36220519445535611814232446305e-9 \\
 -0.062486711605800 & -0.06248670430184261 & 7.30395752607397304088913068161e-9 \\
 0.055642809404103 & 0.05564280387866634 & 5.52543659403385396225617486478e-9 \\
 -0.039269392294065 & -0.03926938964799489 & 2.64607041571345634092514303076e-9 \\
 0.020536072745919 & 0.02053607340972092 & 6.6380213458094874544839647086e-10 \\
 -0.003909743322002 & -0.003909747105156338 & 3.78315400388536404476621594287e-9 \\
 -0.008402891800985 & -0.008402885591472957 & 6.20951202781568861294768233511e-9 \\
 0.015821326011614 & 0.01582131841142720 & 7.60018696843514713716248798368e-9 \\
 -0.018783879401313 & -0.01878387161252608 & 7.78878672122380923843558258608e-9 \\
 0.018255822118049 & 0.01825581533718232 & 6.78086685523181913213714060255e-9 \\
\bottomrule
\caption{List for first coefficients for $\epsilon=1,\;k=3$ and $N=300$}
\end{longtable}

\begin{longtable}{S S S}
\toprule
{Exact} & {Approximated} & {Error} \\
\midrule
\endfirsthead
\toprule
{Exact} & {Approximated} & {Error} \\
\midrule
\endhead
 0.869636929741895 & 0.8696369289154984 & 8.2639632075386751609831267644e-10 \\
 -0.407087815042316 & -0.4070878182163843 & 3.17406833048239917674573473779e-9 \\
 0.216933567694858 & 0.2169335731861694 & 5.49131156283211717998949818922e-9 \\
 -0.106529805425555 & -0.1065298124650571 & 7.0395023831306621215217827201e-9 \\
 0.037935605236408 & 0.03793561326583886 & 8.02943058645432012476782094713e-9 \\
 0.004829744421927 & 0.004829735862776746 & 8.5591504015725216596836599567e-9 \\
 -0.030577449416043 & -0.03057744072603654 & 8.69000598688228879827613952092e-9 \\
 0.044829330540936 & 0.04482932207262626 & 8.46831012375815079851227886724e-9 \\
 -0.051286542403170 & -0.05128653446933563 & 7.93383473531969685753002242774e-9 \\
 0.052521857879731 & 0.05252185075598739 & 7.12374392481443698801992275987e-9 \\
 -0.050357429766389 & -0.05035742369181411 & 6.07457455329548970130946005924e-9 \\
 0.046093071017645 & 0.04609306619440154 & 4.82324341246099166535007361203e-9 \\
 -0.040654544697941 & -0.04065454129043179 & 3.40750960332515101939776703569e-9 \\
 0.034694979501383 & 0.03469497763528018 & 1.86610308011257206446404541296e-9 \\
 -0.028666813694137 & -0.02866681345550335 & 2.3863386658590941315822979567e-10 \\
 0.022874168953499 & 0.02287417038814894 & 1.43465019694225070375611051379e-9 \\
\bottomrule
\caption{List for coefficients for $\epsilon=2,\; k=2$ and $N=300$}
\end{longtable}

As mentioned, we computed also the approximate norm 
squared of the exact coefficients and the approximated ones for different values of the parameters $\epsilon, \; k$ and $N$. 
In the following table we show the error estimates obtained by taking the absolute value of the difference of the exact and approximated norms
(taking the square root of the approximated norm squared). 
The numbers $0 \; \pm \#$ is what Mathematica interprets as zero up to some numerical accuracy $\#$, 
in other words, the true value, due to numerical errors, could be anywhere in the range $\pm \;\#$.
The values for $k=2, \epsilon=1$ seem counter-intuitive since higher $N$ should generally improve accuracy.
Since the numbers are extremely small, the reported decreasing precision at higher $N$ 
appears to be due to numerical noise and thus 
does not indicate a loss of accuracy but rather that the error is so small that it is indistinguishable from zero at higher $N$.
This is crossed-checked for $k=3, \epsilon=1$ where the error estimate is clearly decreasing and the numbers shown for 
$N=100, 150, 200$ are very small but not zero up to a given numerical accuracy.

\begin{longtable}{c c c c}
\toprule
{k} & {$\epsilon$} & {N} & {Norm error} \\
\midrule
\endfirsthead
\toprule
{k} & {$\epsilon$} & {N} & {Norm error} \\
\midrule
\endhead
 2 & 2 & 300 & 0 $\pm$ 4.140505204929362$\;\times\;10^{-10}$ \\
 \hline \\
 2 & 1 & 100 & 0 $\pm$ 3.4462107345939413$\;\times\;10^{-10}$\\
 2 & 1 & 150 & 0 $\pm$ 2.0612713165980796$\;\times\;10^{-8}$ \\
 2 & 1 & 200 & 0 $\pm$ 2.843109927263203$\;\times\;10^{-7}$ \\
 \hline \\
 3 & 1 & 100 & 3.929609116040976$\;\times\;10^{-7}$ \\
 3 & 1 & 150 & 2.162275954188642$\;\times\;10^{-8}$ \\
 3 & 1 & 200 & 6.088486375407583$\;\times\;10^{-10}$ \\
 3 & 1 & 250 & 0 $\pm$ 2.163659196252404$\;\times\;10^{-10}$ \\
 3 & 1 & 300 & 0 $\pm$ 7.853411815031981$\;\times\;10^{-10}$ \\
\bottomrule
\caption{Convergence of the norm error for various parameters $k$, $\epsilon$, and $N$.}
\end{longtable}

\section{Raleigh-Ritz method versus perturbation theory}
\label{s4}

Finally, we will also study how the RRM performs in comparison with the stationary perturbation (PT) theory of 
point spectra. The latter depends on a perturbative expansion in terms of a dimension free 
parameter $\delta$ which is assumed to be well below unity in order that the expansion
has a chance to converge. In our case we identify that parameter as $\delta=[l/L]^{2k}$, that is,
$\delta=\epsilon^{2k}$ in terms of the parameter $\epsilon=l/L$ used in the previous section. Note 
that even for $\epsilon=\frac{1}{2}$ and the lowest possible $k=2$ of interest we have $\delta=0.0625$
so that the assumption on the size of $\delta$ is met even for $\epsilon$ below but not too close to unity.
In what follows we restrict to the case $k=2,\sigma=1$ and such $\epsilon$.

According to the PT procedure, we split the Hamiltonian 
as 
\be \label{2.20}
H=H_0+U, \; H_0=\frac{p^2}{2m}+V_0,\; V_0=\frac{\hbar^2}{2m\; l^2}[\frac{q^2}{l^2}+1]
\ee
thereby collecting all $\epsilon$ dependent terms in (\ref{2.15}) in $U=V_b-V_0$ considered as
a perturbation. The Hamiltonian $H_0$ has the above Hermite functions as eigenstates of the 
unperturbed Hamiltonian. 
For this specific model, the Hamiltonian is parametrised in terms of $U=\delta\; H_1$ i.e.
\begin{align}
\label{Hamiltonian}
H(\epsilon)=H_0+\delta H_1,\;\;\;\; H_0=\frac{1}{2}(-[\frac{d}{dy}]^2+1+y^2),\;\;\;\; 
H_1=\frac{-3y^2-2y^4}{2}+\frac{\delta}{2}\;y^6.
\end{align}
We will compute the ground state and its energy up to second order in perturbation theory,
thus one computes  
$E_n(\delta)=E_n^{(0)}+\delta \;E_n^{(1)} +\delta^2\; E_n^{(2)}$
and $e_n(\delta)=e_n^{(0)}+\delta \; e_n^{(1)}+\delta^2\;e_n^{(2)}$ where the upper index corresponds to the perturbation order
and then specialises to $n=0$.
The formulas for $E_n^{(k)}$ and $e_n^{(k)}$ are standard and can be found in any quantum mechanics text book (see e.g. \cite{1,2}). 
For the lower bound energy the formula is
\begin{align}
\label{pertenergies}
E_n(\delta)=E_n^{(0)}+\delta\braket{e^{(0)}_n,H_1\; e^{(0)}_n}+\delta^2 \sum_{l\neq n}\frac{|\braket{e^{(0)}_l,H_1\; e^{(0)}_n}|^2}{E^{(0)}_n-E^{(0)}_l},
\end{align}
here $E^{(0)}_n$ corresponds to the nth- energy eigenvalue of the unperturbed Hamiltonian $H_0$, while $e^{(0)}_n$ is its corresponding 
(normalised) eigenstate. This formula apples because the unperturbed energy levels $E_n^{(0)}$ are non-degenerate 
so that all fractions that appear are well defined, in fact the denominator is bounded from below in modulus by 
unity. For the eigenstates one has  
\begin{align}
\label{pertstates}
e_n(\delta)&=e_n^{(0)}+\delta\sum_{l\neq n}\frac{\braket{e^{(0)}_l,H_1\; e^{(0)}_n}}{E^{(0)}_n-E^{(0)}_l} e^{(0)}_l
+\delta^2 \sum_{\substack{l\neq n \\
m\neq n}}\frac{\braket{e^{(0)}_l,H_1\; e^{(0)}_m}\braket{e^{(0)}_m,H_1\; e^{(0)}_n}}{(E^{(0)}_n-E^{(0)}_l)(E^{(0)}_n-E^{(0)}_m)}e^{(0)}_l  
- \delta^2\sum_{l\neq n}E^{(1)}_n\frac{\braket{e^{(0)}_l,H_1\; e^{(0)}_n}}{(E^{(0)}_n-E^{(0)}_l)^2} e^{(0)}_l
\end{align}
We chose to work with creation and annihilation operators instead of Hermite functions.
The inner products in \cref{pertenergies,pertstates} are fairly easy to compute by hand, however,
they are tedious particularly when we need to expand $(a+a^\dagger)^6$. 
Since anyway we are interested in the numerical data, we can directly compute the equations in \texttt{Mathematica}. 
Once again we used the \texttt{cuantica} and \texttt{versora} packages and computed the matrix elements in \cref{pertenergies,pertstates} following the code shown in \cref{a2}.

In the following table we show the values of the ground state energies.
The first column corresponds to the truncated Hilbert space dimension which we here implemented 
also within PT by hand, for a better comparison between the two methods, 
by restricting in the above formulae the sums over $k,m\not=n=0$ to the range 
$\{1,2,,..,N\}$.   
The second is the approximated value using stationary PT at second order together with that 
truncation and the third using the RRM.
For $N=25$ both methods produce close results far away from the exact ground state energy eigenvalue which 
here is zero by our whole construction. 
For N=150, the perturbative result for the energy eigenvalues remains unchanged 
as compared to $N=25$ because 
whenever $H_1$ is a polynomial of finite degree $M$ in annihilation and creation operators the 
matrix elements $<e^{(0)}_l,\;H_1\; e^{(0)}_0>$ automatically vanish for $l>M$. Thus for $k=2$ 
truncating above $N=6$ no longer influences the result of PT. On the other hand, the  RRM value 
for $N=150$ becomes extremely small thus very close to the exact eigenvalue.
demonstrating the fast convergence of the RRM results as $N$ increases. In order to improve the 
results of PT one would therefore would need to increase the perturbative order rather 
than the truncation dimension. 
\begin{longtable}{ccc} 
 \toprule
 N & Perturbation & RRM \\
 \midrule
 25 & 0.981603831246387 & 0.9815159620090416 \\
 125 & 0.981603831246387 & 2.7330987307456215$\times 10^{-15}$ \\
 \bottomrule
 \caption{Approximated ground state energies for different truncation dimensions $N$ and $\delta=0.011$}.
\end{longtable}
A similar procedure can be applied to determine the perturbed ground states.  
We set \( N = 125 \) and computed \( e_0 \) using both PT and RRM. Note that here the truncation 
is slightly more influential within PT than for the energy eigenvalues because in the above formula for the perturbed eigenstates to second 
order we find a sum over $l,m\not=0$ and a product of matrix elements of $H_1$ between the unperturbed eigenstates
of level $l,m$ and $m,0$ respectively. Thus we find contributions only for $m\le 6$ and $l\le 12$, accordingly 
the truncation plays no role beyond $N=12$. 
Since the eigenstates are normalisable, we numerically estimated the error by evaluating the deviation of the truncated norm
defined by the corresponding Fourier coefficients from unity.
For PT this error estimate was 
$0.00021411438895824197$, while for the RRM we got $5.568878691519785\times 10^{-13}$.
The analysis is completely analogous to the one performed for the ground energy.
Thus once again, fast convergence is achieved using RRM by increasing $N$ 
while for PT we would need to go instead higher orders in $\delta$.

The expressions for higher orders in perturbation theory are far from trivial, 
moreover the perturbation accuracy is very sensitive of $\delta$.
On the other hand, RRM is in this sense non-perturbative and shows fast convergence for several values of 
$\delta$ even above unity. The price to pay is that RRM  
is computationally more expensive since one requires to compute matrices of increasing dimension.

\section{Conclusion}
\label{s5}

In the present work we have used ideas from supersymmetric quantum mechanics in oder 
to engineer a class of Hamiltonians which i. have an explicitly known ground state and 
ii. model the physically interesting situation of a potential that allows for spontaneous 
symmetry breaking for a certain choice of its parameters. That is to say, if one works with a position 
operator shifted by one of the minima of the potential (of which there is an even number) then
that position operator has a condensate  (i.e. 
a non vanishing expectation value) with respect to the exact ground state given by that minimum.
Such shifted position operators are of interest because one may use the 
position operator shifted by the potential minimum to build corresponding annihilation and creation 
operators in a Fock quantisation of the Hamiltonian. While a mere rewriting in quantum mechanics,
in QFT such different Fock quantisations do make a difference 
as different Fock representations are generically not in the same unitarity class of the 
representations of the canonical commutation relations.

The availability of the exact ground state allowed us to rigorously assess the accuracy of the RRM by directly computing the error between the exact and approximate results.  
Convergence of the method depends on the parameters $\epsilon$, $k$, and $N$.  
For fixed values of $\epsilon$ and $k$, increasing $N$ systematically reduces the error.  
As shown in the tables, 
convergence is significantly faster for small values of $\epsilon$ and $k$, 
allowing accurate results to be obtained with relatively small matrices, 
an advantage in terms of computational efficiency.  
In our study, we focused on the parameter range $0 < \epsilon < 5$ and $1 < k < 5$.  
This choice was due to memory constraints, limits in numerical precision, and general machine capabilities, 
as we observed that meaningful convergence often required $N \gg 300$, which quickly became computationally demanding.

Table 13 highlights the effectiveness of the RRM: 
for several combinations of the parameters $\epsilon$, $k$, and $N$, the numerical error is zero up to machine precision. 
In such settings, the method demonstrates remarkable performance. 
In contrast, stationary perturbation theory performs rather poorly, especially for systems with strong interactions, as is also observed in the case of the anharmonic oscillator~\cite{5}. 
The RRM is non-perturbative and remains applicable even in strongly coupled regimes. 

The methods presented in this work can be easily extended to the case of arbitrary polynomial 
superpotentials which are engineered from a given normalisable state which is given as the exponential 
of a polynomial bounded from above and unbounded from below.\\  
\\ 
\\       
{\bf Acknowledgements}\\
M.R.-Z. thanks Jonas Neuser for useful discussions on the topic and acknowledges the financial support provided by the Deutsche Akademische Austauschdienst e. V.\\

\begin{appendix}

\section{Mathematica codes}
\label{a2}

\subsection{Code for Raleigh-Ritz method}

To load the custom packages we us the \texttt{Needs} command. The packages require to set the coordinate's system. 
\begin{Verbatim}
Needs["cuantica`","c:directory path"];
Needs["versora`","c:directory path"];
SetCoordinates[e][Cartesian[{x,y,z}], Cylindrical[{rhc,thc}],Spherical[{r,th,ph}]];
\end{Verbatim}
To define the discrete orthonormal basis;
\begin{verbatim}
SetDiscreteOB[Ket[HO, {0, 1}]];
\end{verbatim}
We define the creation and annihilation operators. 
The first four lines define the standard annihilation and creation operators using the orthonormal basis previously defined. 
The command \texttt{SetLinearOperator} tells the packages that \texttt{Annihilation} and \texttt{Creation} are linear operators and
enables automated simplification and symbolic manipulation of operator expressions.
\begin{verbatim}
Annihilation[ Ket[HO, m_]] := Sqrt[m] Ket[HO, m - 1];
Annihilation[Ket[HO, 0]] = 0;
Creation[ Ket[HO, m_]] := Sqrt[m + 1] Ket[HO, m + 1];
Creation[0] = 0;
SetLinearOperator[Creation, Ket];
SetLinearOperator[Annihilation, Ket];
\end{verbatim}
Next we define position and momentum operators in terms of \texttt{Creation} and \texttt{Annihilation}. 
\texttt{HIP} is the built-in Hilbert space inner product. 
\texttt{X[X[ Ket[HO, k] ]]} corresponds to the quadratic operator $x^2$ acting on the ket.
Thus lines 3 and 4 are the expectation values for the operators $X^2$ and $P^2$ respectively
\begin{verbatim}
X = 1/Sqrt[2] (Annihilation[#] + Creation[#]) &;
P = 1/(I Sqrt[2]) (Annihilation[#] - Creation[#]) &;		
HIP[Bra[HO, k], X[X[ Ket[HO, k]]]] //Expand;
HIP[Bra[HO, k], P[P[ Ket[HO, k]]]] // Expand;
\end{verbatim}
To construct the Hamiltonian matrix $H^N_{m,n}$. 
$k$ parametrises the degree of the polynomials in \cref{2.15} while $\epsilon$ is the coupling parameter and
$N$ is the dimension of the matrix. 
\texttt{comp1, comp2, comp3} account for $x^2,\; x^{2k-2}, \; x^{4k-2}$ in \cref{2.15} respectively. 
Lines 7th to 9th correspond to the operators  $x^2,\; x^{2k-2}, \; x^{4k-2}$ acting on the basis $b_n$.
The last two lines are short cuts of the $\epsilon$-dependent coupling parameters depicted in \cref{2.15}
\begin{verbatim}
k = 3;
N = 5;
\[Epsilon] = 2;
comp1 = Apply[Composition, ConstantArray[X, 2 k]];
comp2 = Apply[Composition, ConstantArray[X, 2 k - 2]];
comp3 = Apply[Composition, ConstantArray[X, 4 k - 2]];
comp1[Ket[HO, n]] // Expand;
comp2[Ket[HO, n]] // Expand;
comp3[Ket[HO, n]] // Expand;
eps2k = \[Epsilon]^(2 k);
eps4k = \[Epsilon]^(4 k);
\end{verbatim}
The Hamiltonian matrix is obtained by applying the built-in inner product command \texttt{HIP}. 
Since the matrix contains many zero entries, we represent it as a \texttt{SparseArray}, 
this improves computational efficiency in both memory usage and numeric operations.
\begin{verbatim}
matk = SparseArray[Table[N[(n + 1) KroneckerDelta[m, n] - 
((2 k - 1)/2) eps2k HIP[ Bra[HO, m], comp2[Ket[HO, n]]] - 
eps2k HIP[Bra[HO, m], comp1[Ket[HO, n]]] + 
1/2 eps4k HIP[Bra[HO, m], comp3[Ket[HO, n]]]], {m, 0, M}, {n, 0,M}]];
\end{verbatim}
Finally, we are only interested in the lowest energy eigenvalue and its associated eigenvector. 
This can be extracted from the matrix using the following code
\begin{verbatim}
{smallestEigenvalue, smallestEigenvector} = Eigensystem[matk, -1]
\end{verbatim}

\subsection{Code for Perturbation Theory}

We use the same code as before up to the construction of the position and momentum operators. 
To generate the matrix elements of the free Hamiltonian and obtain the ground-state energy
\begin{verbatim}
groundenergy=Table[N[(n + 1) KroneckerDelta[m, n]], {m, 0, M},{n, 0, M}];
{eigenvalues, eigenvectors} = Eigensystem[groundenergy];
\end{verbatim}
The interacting Hamiltonian has polynomials of degree 2, 4 and 6. They are generated by applying the \texttt{Composition} command to the position operators 
\begin{verbatim}
X2 = Composition[X, X];
X4 = Composition[X, X, X, X];
X6 = Composition[X, X, X, X, X, X];
\end{verbatim}
The terms \(E1:=\braket{e^{(0)}_n,H_1\; e^{(0)}_n}\) and \( E2:=\sum_{l\neq n}\frac{|\braket{e^{(0)}_l,H_1\; e^{(0)}_n}|^2}{-l}\) are obtained by constructing the Hamiltonian $H_1$ using \texttt{X2, X4, X6} and then applying the Hilbert space inner product with the \texttt{HIP} command. 
The last line corresponds to the ground state energy up to second order in PT \cref{pertenergies}.
\begin{verbatim}
N=125;
\[delta]=(0.33)^4 // N;

E1 = -3/2 HIP[Bra[HO, 0], X2[ Ket[HO, 0 ]]]
- HIP[Bra[HO, 0], X4[Ket[HO, 0 ]]] + \[delta]/2 HIP[Bra[HO,0], X6[ Ket[HO, 0 ]]] // N;

E2 = Sum[Abs[-3/2 HIP[Bra[HO, l], X2[Ket[HO, 0 ]]] - HIP[Bra[HO, l], X4[ Ket[HO, 0 ]]] +
\[delta]/ 2 HIP[Bra[HO, l], X6[ Ket[HO, 0 ]]]]^2/-l, {l, 1, N}] // N;

E = 1 + \[delta] E1 + \[delta]^2 E2 // N;
\end{verbatim}
In the following we show only the code for the linear term in $\delta$ of \cref{pertstates}.
The codes of the other terms are obtained in a similar fashion.
In the first line we computed the values $\frac{\braket{e^{(0)}_l,H_1\; e^{(0)}_n}}{-l}$ for \(l\neq 0\) and for Hilbert space dimension $N$. 
In the second line we multiplied these values with $e^{(0)}_l$ using the \texttt{MapThread} command. 
\texttt{Rest[Reverse[eigenvectors]]} generates the $e^{(0)}_l$ by taking all the eigenvectors of the matrix \texttt{groundenergy} 
except for the ground state.
In the last line we applied the sum.

\begin{verbatim}
sum1 = Table[1/-l(-3/2 HIP[Bra[HO, l],X2[ Ket[HO, 0 ]]] - HIP[Bra[HO, l], X4[ Ket[HO, 0 ]]] +
\[Lambda]/2 HIP[Bra[HO, l], X6[ Ket[HO, 0]]]), {l, 1, N}] // N;
sumvec1 = MapThread[#1*#2 &,{sum1,Rest[Reverse[eigenvectors]]}];
vec1 = Total[sumvec1];
\end{verbatim}

\end{appendix}



\begin{thebibliography}{99}

\parskip -5pt

\bibitem{1} F. Schwabl, Quantenmechanik: Eine Einf\"uhrung. Springer-Verlag, Berlin, 2007.

\bibitem{2} A. Messiah. Quantum Mechanics. Vol. 1, 2. Dover Publications Inc., Dover, 2014. 

\bibitem{3} T. Kato. Perturbation Theory for Linear operators. Springer-Verlag, Berlin.

\bibitem{4} M. Reed, B. Methods of modern mathematical physics. Vol. 4. Academic Press, 1978.

\bibitem{5} C.M. Bender, T.T. Wu. Phys. Rev. {\bf 184} 184.\\
B. Simon. Coupling constant analyticity for the anharmonic oscillator.\\
j. J. Loeffl, A. Martin, B. Simon, A.S. Wightman. Pade approximants and the anharmonic oscillator.
Phys. Lett. {B30} (1969) 656. 

\bibitem{6} M. Reed, B. Methods of modern mathematical physics. Vol. 2. Academic Press, 1978.


\bibitem{7} H. Cycon, R. Froese, W. Kirsch, B. Simon: Schr\"odinger Operators with
Applications to Quantum Mechanics and Global Geometry. Texts Monographs Phys. Springer, 1987

\bibitem{8} I. S. Gradshteyn, I. M. Ryzhik: Table of Integral, Series and Products. 8th edition. Academic Press 2014

\bibitem{9} NIST Digital Library of Mathematical Functions. https://dlmf.nist.gov/

\bibitem{10} R.B. Paris, D. Kaminski: Asymptotics and Mellin-Barnes Integrals. 
Encyclopedia of Mathematics and its Applications. Cambridge University Press; 2001:vii-xii.

\bibitem{11} R. Cabrera: Quantum Mechanics (cuantica.m): A Mathematica package for quantum mechanical calculus. 
University of Windsor. Wolfram Library Archive. 2001. https://library.wolfram.com/infocenter/ID/494/

 \end{thebibliography}
\end{document}